\documentclass[10pt,prl,twocolumn,showpacs,superscriptaddress]{revtex4}

\usepackage{graphicx}

\newcommand{\be}{\begin{equation}}
\newcommand{\ee}{\end{equation}}
\newcommand{\bea}{\begin{eqnarray}}
\newcommand{\eea}{\end{eqnarray}}
\newcommand{\down}{\downarrow}
\newcommand{\up}{\uparrow}

\newcommand{\f}{\frac}

\begin{document}

\title{Entanglement and Spontaneous Symmetry Breaking in Quantum Spin Models}

\author{Olav F. Sylju{\aa}sen}
\email{sylju@nordita.dk}
\affiliation{NORDITA, Blegdamsvej 17, DK-2100 Copenhagen {\O}, Denmark}


\date{\today}

\pacs{03.67.Mn, 75.10.Jm}
\preprint{NORDITA-2003-44 CM}

\begin{abstract}
It is shown that spontaneous symmetry breaking does not modify the ground-state entanglement of two spins, as defined by the concurrence, in the XXZ- and the transverse field Ising-chain. Correlation function inequalities, valid in any dimensions for these models, are presented outlining the regimes where entanglement is unaffected by spontaneous symmetry breaking.  
\end{abstract}

\maketitle

Entanglement is a property of a quantum state shared between 
two or more parties. It is defined with the aim of capturing the 
essential quantum non-locality encoded in the state. 
While for a long time interests in entanglement stemmed from the opportunity to understand fundamental concepts in quantum mechanics such as the EPR-paradox and violation of Bell-inequalities, 
recent interest in entanglement comes from
its use as a {\em resource} for {\em performing tasks} not possible
by classical means. 

From an applied viewpoint it is thus worth quantifying the
degree of entanglement in natural systems, such as solid-state materials. 
A number of studies have been devoted to quantifying such ``natural''
entanglement in states of simple models describing idealized
quantum magnets, such as the XXZ- and transverse field Ising-models\cite{Gunlycke,Arnesen,Bose,Osterloh,Osborne,Glaser}. However the possible alteration of entanglement by spontaneous symmetry breaking (SSB) were not discussed in any of these works, although the need for such a study was mentioned in Ref.~\cite{Osborne}. 
SSB happens invariably in real materials
described by a Hamiltonian possessing a global symmetry, 
thus it is important to investigate whether or not the entanglement
calculated in the symmetric ground state(s) is changed by SSB.
Here we find that entanglement is {\em not} modified by
SSB for the XXZ- and transverse field Ising-chains.

It is a fundamental requirement of entanglement that it cannot
on average be created by mixing classically two quantum states.
Classical mixing is incoherent, and so one should not
gain more entanglement from this mixing than 
what is already encoded in the states that are being mixed.
Thus if we consider the density matrix $\rho = (\rho_+ + \rho_-)/2$
put together by an equal mixture of the broken symmetry states, $\rho_\pm$ 
(not necessarily pure), the entanglement $E(\rho)$ must
satisfy
\be \label{convex}
   E(\rho) \leq \f{1}{2} \left( E(\rho_+) + E(\rho_-) \right) = E(\rho_\pm)
\ee
where we have assumed in the last equality that the entanglement in the 
two different symmetry-broken ground states are equal, the states being related by a global change of basis states.
Thus the entanglement in the broken state cannot be smaller than
in the symmetric state. 

Here we focus on the entanglement of two specific spins (or qubits) in the ground-state of a quantum system. 
The state of two spins $i$ and $j$ in the ground state of a quantum system
is described in terms of the reduced density matrix $\rho_{ij}$ 
obtained by tracing over all spins in the ground state except the two spins $i$ and $j$. Writing it out explicitly in the standard basis $ \{ |\up \up \rangle,|\up \down \rangle,|\down \up \rangle, |\down \down \rangle \}$ one has
\be \label{densitymatrix}
 \rho_{ij} =\left(
 \begin{array}{llll}
   \langle P^\up_i P^\up_j \rangle &
   \langle P^\up_i \sigma^-_j      \rangle &
   \langle \sigma^-_i P^\up_j      \rangle &
   \langle \sigma^-_i \sigma^-_j   \rangle  \\
   \langle P^\up_i \sigma^+_j      \rangle &
   \langle P^\up_i P^\down_j       \rangle &
   \langle \sigma^-_i \sigma^+_j   \rangle &
   \langle \sigma^-_i P^\down_j    \rangle  \\
   \langle \sigma^+_i P^\up_j      \rangle &
   \langle \sigma^+_i \sigma^-_j   \rangle &
   \langle P^\down_i P^\up_j       \rangle &
   \langle P^\down_i\sigma^-_j     \rangle  \\
   \langle \sigma^+_i \sigma^+_j   \rangle &
   \langle \sigma^+_i P^\down_j    \rangle &
   \langle P^\down_i \sigma^+_j    \rangle &
   \langle P^\down_i P^\down_j     \rangle 
\end{array}
\right)
\ee
where $P^\up = \f{1}{2}(1+\sigma^z)$, $P^\down = \f{1}{2}(1-\sigma^z)$
and $\sigma^\pm = \f{1}{2}\left(\sigma^x \pm i \sigma^y \right)$.
The brackets denote ground state expectation values and $\sigma$ are the Pauli matrices.

Recently entanglement between blocks of consecutive spins have been studied using the von Neumann entropy\cite{Vidal} which is the conventional measure of bipartite entanglement in pure states.
There it is essential that the ground-state is pure as the von Neumann entropy is a concave function violating Eq.~(\ref{convex}). This requirement becomes very restrictive when considering the state of just two spins tracing out the rest. Then even though the ground-state of the full system is pure, $\rho_{ij}$ is generally not. Thus a definition of entanglement in mixed states is needed. A reasonable definition is the entanglement of formation, $E_f$ \cite{Ef} which is obtained by decomposing the mixed ensemble into pure states, and summing the entanglement in each of these pure states weighted by their probabilities of occurrence in the mixed ensemble. As there are many possible decompositions into pure states, $E_f$ is taken to be the minimal value gotten by trying all possible decompositions. 
This definition of entanglement clearly satisfies Eq.~(\ref{convex}) as
the minimal decompositions of $\rho_+$ and $\rho_-$ into pure states also
is a decomposition of $\rho$, but not necessarily the minimal one. 
Furthermore $E_f$ in pure states reduces to the conventional definition of entanglement.

The minimization procedure over all possible decompositions is however quite
difficult to handle practically. It is therefore of great value that
there exist a closed-form expression for this minimum value for bipartite
two-level systems. 
This expression is in terms of the concurrence ${\cal C}$\cite{HillWootters}, which is defined in terms
of the spectrum of the matrix $\rho_{ij} \tilde{\rho_{ij}}$ where $\tilde{\rho} = \sigma^y_i \otimes \sigma^y_j {\rho}^* \sigma^y_i \otimes \sigma^y_j$, the time-reversed density matrix. The complex conjugation refers to the standard basis used in Eq.~(\ref{densitymatrix}). Let $\lambda_l$ be the eigenvalues of $\rho \tilde{\rho}$ so that $\lambda_1 \geq \lambda_2 \geq \lambda_3 \geq \lambda_4$. Then the concurrence ${\cal C}$ is
\be \label{concurrence}
    {\cal C} = {\rm max} \left\{ 0, \sqrt{\lambda_1}-\sqrt{\lambda_2} -\sqrt{\lambda_3} -\sqrt{\lambda_4} \right\}
\ee
and the entanglement of formation is $E_f = -x \log_2 x -(1-x) \log_2 (1-x)$,
where $x = 1/2+\sqrt{1-{\cal C}^2}/2$. Because $E_f$ is a monotonous function of ${\cal C}$ with $E_f({\cal C}=0)=0$ and $E_f({\cal C}=1)=1$ we will hereafter for simplicity discuss the concurrence instead of the entanglement of formation.
In Ref.\cite{Uhlmann} it was shown that the concurrence as defined above is the largest convex function reducing to the pure state concurrence, thus ${\cal C}$ satisfies Eq.~(\ref{convex}).

Before discussing mixed ensembles of general states we will first discuss the ensemble obtained by an equal classical mixture of two {\em pure} states with equal concurrences. That is $\rho = (\rho_+ + \rho_-)/2$ and $\rho_\pm = |\alpha_\pm \rangle \langle \alpha_\pm|$. The two states $|\alpha_\pm \rangle$ can be thought of as different realizations of a broken global symmetry. 
Denoting the concurrence $c=|\langle \alpha_+ | \tilde{\alpha}_+ \rangle | = | \langle \alpha_- | \tilde{\alpha}_- \rangle |$, and the overlap $\langle \alpha_+ | \tilde{\alpha_-} \rangle = d$ the square roots of the eigenvalues of $\rho \tilde{\rho}$ are $|c \pm |d||/2$. Thus
\be
   {\cal C}(\rho) = {\rm min} \left\{ c,|d| \right\}.
\ee
So the concurrence of the symmetric mixed state equals the concurrence of the broken symmetry state only when $| \langle \alpha_+ | \tilde{\alpha_+} \rangle | \leq   | \langle \alpha_+ | \tilde{\alpha_-} \rangle |$.
From this it becomes clear that the question of the influence of SSB on concurrence cannot be answered in general terms, one
needs to look at the specifics of the states involved. We will therefore investigate two specific models; the XXZ- and the transverse field Ising-model.

The Hamiltonian for the XXZ-model is
\be
H_{\rm xxz} = \sum_{\langle i,j \rangle} \left\{
              -\left(  \sigma^x_i \sigma^x_j 
	                     + \sigma^y_i \sigma^y_j \right) + 
	      \Delta \sigma^z_i \sigma^z_j \right\},
\ee
where the sum is taken over all nearest neighbor sites on a lattice which for simplicity is taken to be bipartite. The Hamiltonian is real and invariant under a $U(1)$-rotation about the spin z-axis. This symmetry being continuous can only be broken in dimensions higher than one and for $|\Delta |< 1$. A global $\pi$-rotation about the spin x (or y) axis is also a symmetry ($Z_2$). This symmetry can be broken in one dimensional systems when $|\Delta| \geq 1$. Let us first consider the breaking of the $Z_2$-symmetry assuming the $U(1)$-symmetry is unbroken. The
reduced density matrix then reads
\be \label{Z2densitymatrix}
 \rho_{ij} =\left(
 \begin{array}{llll}
    A & 0 & 0 & 0 \\
    0 & B & C & 0 \\
    0 & C & G & 0 \\
    0 & 0 & 0 & D
\end{array}
\right)
\ee  
where the symbols $A \ldots G$ expressed in terms of correlation functions can be read off Eq.~(\ref{densitymatrix}). With a matrix of this form the concurrence is ${\cal C} = 2 \cdot {\rm max} \{0,|C|-\sqrt{AD} \}$ which in terms of correlation functions is
\bea
{\cal C} & = & \f{1}{2} \cdot \rm{max} \left\{ \f{}{} 0 , \right. \\
&  &  \left. |\langle \sigma_i^x \sigma_j^x \rangle 
            + \langle \sigma_i^y \sigma_j^y \rangle | 
		   - 
		   \sqrt{ (1+ \langle \sigma^z_i \sigma^z_j \rangle )^2
		     - \langle \sigma^z_i + \sigma^z_j \rangle^2 }
		   \right\} \nonumber
\eea
The breaking of the $Z_2$ symmetry manifests itself as 
$\langle \sigma^z \rangle \neq 0$. The values of the other correlators
are unchanged by SSB as $\sigma_i^x \sigma_j^x$, $\sigma_i^y \sigma_j^y$, $\sigma_i^z \sigma_j^z$ are all invariant under the $Z_2$ transformation.

For 
$\Delta \leq -1$ SSB does not affect the concurrence as it is zero in both cases, the ground state being either a product state or a sum of product states. Thus the interesting region is the antiferromagnetic Ising region $\Delta \geq 1$ where $\langle \sigma^z_i \rangle = \pm m$, the sign depending on which sublattice spin $i$ is located on. For $i$ and $j$ on different sublattices
, the concurrence is not affected by the symmetry-breaking as then $\langle \sigma^z_i + \sigma^z_j \rangle =0$. For $i$ and $j$ on the same sublattice knowledge about the values of the correlation functions
is needed. To obtain these we have performed Monte Carlo simulations
using the Stochastic Series Expansion technique\cite{SSE} with directed-loop
updates\cite{SS}. The correlation functions were obtained directly
from the loop movement in space-time in analogous manner to that described in Ref.~\cite{WORMS}. In order to simulate the broken symmetry phase a small symmetry breaking staggered field were imposed.  
\begin{figure}
\includegraphics[clip,width=8cm]{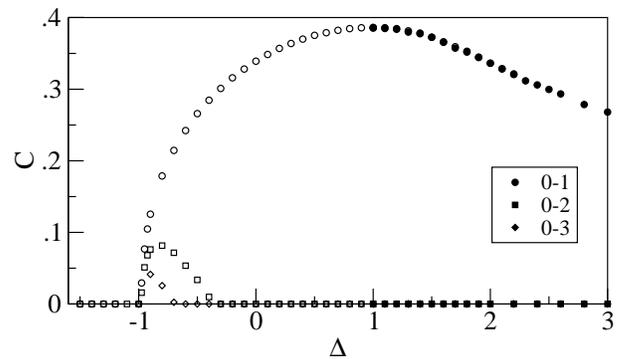}
\caption{The concurrence as function of $\Delta$ for different separations between sites $i$ and $j$ for a 128 site XXZ-chain with(filled symbols, only for $\Delta \geq 1$) and without (open symbols) a symmetry breaking staggered field $h_z=0.001$. The inverse temperature is $1/T =80$. For $\Delta \geq 1$ the open and filled symbols lie on top of each other.}
\label{xxzChain}
\end{figure}   
The results can be seen in Fig.~\ref{xxzChain} which shows the concurrence as a
 function of $\Delta$ for different separations between $i$ and $j$ both with and without a symmetry-breaking staggered field along the spin z-axis. As can be seen the symmetry-breaking field has no effect on the concurrence.
For $i$ and $j$ nearest neighbors the concurrence reaches a maximum of $0.386$ at $\Delta=1$ in agreement with Ref.~\cite{Oconnor}, while for bigger
distances between $i$ and $j$ the concurrence drops rapidly to zero long before
$\Delta$ reaches 1. 

In dimensions greater than one the $U(1)$-symmetry is broken when $ | \Delta |< 1$. If the $Z_2$-symmetry is obeyed, the density matrix takes the form
\be \label{xxzU1densitymatrix}
 \rho_{ij} =\left(
 \begin{array}{llll}
    A & a & a & f \\
    a & B & C & a \\
    a & C & B & a \\
    f & a & a & A
\end{array}
\right)
\ee
where $a=\langle \sigma^x_i \rangle/4= \langle \sigma^x_j \rangle/4$ and $f=\langle \sigma^x_i \sigma^x_j -  \sigma^y_i \sigma^y_j \rangle /4$ are the only values affected by the breaking of the $U(1)$-symmetry. (The case when a magnetic field is present along the spin-z axis is discussed at the very end of this article.) 
The square roots of the eigenvalues of $\rho \tilde{\rho}$ written in terms 
of the correlation functions are in this case
\bea
u_{\pm}    & = & \f{1}{4} \left| \sqrt{ \left( 1+ \langle \sigma^x_i \sigma^x_j \rangle \right)^2
                    -4 \langle \sigma^x_i \rangle^2}
        \pm | \langle \sigma^y_i \sigma^y_j \rangle
	- \langle \sigma^z_i \sigma^z_j \rangle  | \right|  , \nonumber \\
v_{\pm} & = &
	\f{1}{4} \left| 1 - \langle \sigma^x_i \sigma^x_j \rangle
	\pm \left( \langle \sigma^y_i \sigma^y_j \rangle
	+\langle \sigma^z_i \sigma^z_j \rangle \right) \right|.
	\label{U1eigen}
\eea
In the symmetric state $\langle \sigma^x \rangle = 0$ and
$\langle \sigma^x_i \sigma^x_j \rangle =\langle \sigma^y_i \sigma^y_j \rangle$.
The condition that $\rho_{ij}$ has non-negative eigenvalues implies that whenever the concurrence is non-zero in the symmetric state, $u_+$ is the largest of the square roots of the eigenvalues.
Now assume that the symmetry is broken in a continuous manner and that the concurrence is non-zero in the symmetric state, then $u_+$ remains the largest until another eigenvalue takes over that role, at which point the concurrence becomes zero. However, if this were to happen the fundamental criterion Eq.~(\ref{convex}) would be violated\cite{footnote}. Thus we conclude that $u_+$ remains the largest even when the symmetry is broken.

The concurrence in the symmetry broken state then takes the $U(1)$-invariant form
\be \label{U1conc}
{\cal C} =  \f{1}{2} \left(
          \langle \sigma^x_i \sigma^x_j \rangle 
                +\langle \sigma^y_i \sigma^y_j \rangle
		-\langle \sigma^z_i \sigma^z_j \rangle -1 \right). 
\ee
iff $\langle \sigma^y_i \sigma^y_j \rangle + \langle \sigma^z_i \sigma^z_j \rangle > \langle \sigma^x_i \sigma^x_j \rangle - 1$ and $\langle \sigma^y_i \sigma^y_j \rangle > \langle \sigma^z_i \sigma^z_j \rangle$. If these inequalities are not satisfied, the concurrence will take on different $U(1)$-{\em non}-invariant expressions in the broken state. However, of these only (\ref{U1conc}) reduces smoothly 
(it is constant) to the non-zero value gotten at $a,f\to 0$, 
thus we conclude that
the expression (\ref{U1conc}) is always valid in a regime close to the symmetric state provided the concurrence in the symmetric state is non-zero. The results for the case of antiferromagnetic interactions in the spin-xy-plane can be obtained by changing signs on $\sigma^x$ and $\sigma^y$ on one sublattice in the ferromagnetic results described here.

\begin{figure}
\includegraphics[clip,width=8cm]{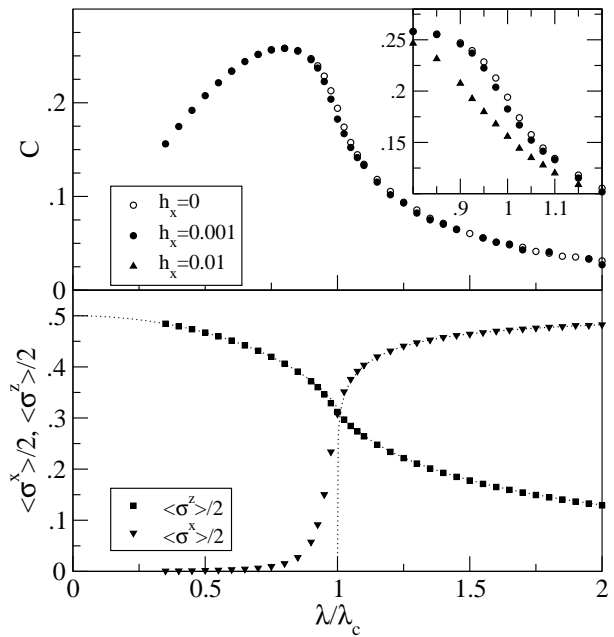}
\caption{Ising chain with 256 sites in a transverse field with and without a symmetry breaking field. The inverse temperature is $1/T=80$. The top panel show the nearest-neighbor concurrence ${\cal C}$ calculated from the measured nearest neighbor density matrix vs. $\lambda$ for different values of the symmetry breaking field. The open symbols means $h_x=0$ while the closed symbols is for $h_x=0.001$. The inset in the top panel shows a blow-up of the region close to the critical point $\lambda=lambda_c$ as well as data for $h_x=0.01$ (filled triangles). 
The bottom panel show the transverse and longitudinal magnetization for the case $h_x=0.001$ as functions of $\lambda$.
The dotted lines are exact results at $T=0$ for zero symmetry breaking field\cite{Lieb}, the value of $\langle \sigma^x \rangle$ is gotten from $\langle \sigma^x_i \sigma^x_j \rangle \asymp \langle \sigma^x \rangle^2$ for $|i-j| \to \infty$. }
\label{traIsing_plot}
\end{figure}

The Hamiltonian for the transverse field Ising model is
\be
 H = - \sum_{\langle i,j \rangle} 
                          \sigma^x_i \sigma^x_{j} 
			 +h_z \sum_i \sigma^z_i
\ee
This model has less symmetries than the XXZ-model. In fact the only
symmetry is a global $\pi$-rotation about the spin $z$-axis. This
symmetry being discrete is broken for $\lambda \equiv 1/(2h_z) > \lambda_c$ in any dimensions. 
To gain insight we have performed Monte Carlo simulations both with and
without a symmetry breaking field $h_x$ along the spin-x axis, see Fig.~\ref{traIsing_plot}.  The results without a symmetry breaking field is in good agreement with the results obtained in Refs.~\cite{Osterloh,Osborne}.
From the upper panel in Fig.~\ref{traIsing_plot} it is seen that the effect of the symmetry breaking magnetic field is to slightly lower the concurrence in the critical region around $\lambda_c$. This would seem to contradict Eq.~(\ref{convex}), however from the inset in the upper panel we see that this effect diminishes as $h_x \to 0$, thus we interpret the lowering as an effect of the small but {\em finite} $h_x$, which will change the symmetry broken states {\em themselves} whenever the gap to excitations is small as is the case close to the critical point. The deviation from the ideal situation when $h_x$ just acts as a symmetry breaking field can also be seen in the lower panel where the magnetization is seen to deviate slightly from the exact result close to $\lambda_c$. We conclude that the Monte Carlo results indicate that SSB does not alter the nearest-neighbor concurrence in the transverse field Ising-chain.

To explain this in more details consider   
the density matrix for two spins in the transverse field Ising model
\be \label{Isingdensitymatrix}
 \rho_{ij} =\left(
 \begin{array}{llll}
    A & a & a & F \\
    a & B & C & b \\
    a & C & B & b \\
    F & b & b & D
\end{array}
\right),
\ee  
where the symmetry-breaking manifests itself in non-zero values of $a$ and $b$ reflecting the non-zero values of $\langle \sigma^\pm \rangle$ and $ \langle \sigma^\pm \sigma^z \rangle$.
For the symmetric state where $a=b=0$, the square root of the eigenvalues of $\rho \tilde{\rho}$ are
$B \pm C$ and $\sqrt{AD} \pm F$, thus the concurrence is
${\cal C} = 2 \cdot {\rm max} \left\{ 0,|C|-\sqrt{AD}, |F|-B \right\}$.
For general values of $a,b$ the eigenvalues of $\rho \tilde{ \rho}$ is gotten
by solving the determinantal condition which amounts to solving a cubic polynomial equation as the eigenvalue $|B-C|^2$ factorizes out.
Instead of solving for the eigenvalues we formulate a condition the coefficients $g_0$,$g_1$ and $g_2$ of this cubic polynomial must satisfy in order for the concurrence to stay constant. Let $x^2$, $y^2$ and $z^2$ be the roots of the cubic polynomial in increasing order. Then
the equations relating these eigenvalues to the coefficients are
\bea
   g_0 & = & (xyz)^2, \nonumber \\
   g_1 & = & x^2 y^2 + x^2 z^2 + y^2 z^2, \\
   g_2 & = & x^2+y^2+z^2. \nonumber 
\eea
The condition that the concurrence stays constant is
\be \label{kappa}
   z-(x+y) = \kappa, 
\ee
where $\kappa$ is a constant equal to the $a=b=0$ value. Squaring Eq.~(\ref{kappa}), rearranging, and squaring again one
finds that the roots can be completely eliminated
in favor of the coefficients, thus
\be \label{inv}
  2 \kappa \sqrt{g_0} = \f{1}{4} \left( \kappa^2 -g_2 \right)^2-g_1.
\ee
If this equation holds for $\kappa$ constant 
as the coefficients $g$ are varied, the concurrence is unaffected by
SSB. 
Written out 
\bea
  g_0 & = & [\alpha^2-4\gamma \delta] \beta -4\mu \nu \alpha -4\mu^2 \delta +4\nu^2 \gamma  \nonumber \\
  g_1 & = & \alpha^2 +2\alpha \beta -4\mu \nu -4\gamma \delta \\  
  g_2 & = & 2\alpha + \beta  \nonumber 
\eea
where $\alpha = F^2+AD-2ab$, $\beta = (B+C)^2-4ab$, $\gamma =DF-b^2$, 
$\delta=AF-a^2$, $\mu=aD-b(B+C-F)$ and $\nu=a(B+C-F)-bA$.
Squaring Eq.~(\ref{inv}) and inserting the expressions for the $g$'s 
one can check that Eq.~(\ref{inv}) holds for
nonzero $a$ and $b$ iff $\sqrt{AD}+F$ is the biggest of the square roots of the eigenvalues; i.e. $\kappa = 2F -(B+C)$.   
In terms of correlation functions this condition is
\be
   \sqrt{(1+\langle \sigma_i^z \sigma_j^z \rangle )^2 - 4 \langle \sigma_i^z \rangle^2} + \langle \sigma_i^z \sigma_j^z \rangle -1 > 2 \langle \sigma_i^y \sigma_j^y \rangle.
\ee  
We have checked numerically that this inequality is satisfied for non-zero values of $h_z$ for the nearest neighbor correlators in the transverse field Ising chain with no symmetry-breaking field, thus SSB does not alter nearest-neighbor concurrence in this case. 

The reduced density matrix for the XXZ-model in the presence of an external magnetic field along the spin z-axis takes the same form as Eq.~(\ref{Isingdensitymatrix}) with the difference that $F=f$ is not invariant under the $U(1)$-symmetry rotation about the spin z-axis. In the $U(1)$-symmetric case, $a=b=f=0$ 
and so whenever concurrence is non-zero $|B+C|^2$ is the largest eigenvalue, implying that
$\kappa = B+C-2\sqrt{AD}$. However with this value of $\kappa$ Eq.~(\ref{inv}) is not satisfied for non-zero values of $a,b$ and $f$. Thus we conclude that
in this case SSB is generally accompanied by a change in the concurrence. 
 
While we have given conditions for when the concurrence is not changed by SSB in some specific models it would be nice if more general statements can be made. It is reasonable to expect that these will involve statements about the system's behavior under time-reversal.

\begin{acknowledgments}
The author acknowledges helpful discussions with Alan Luther.
\end{acknowledgments}


\begin{thebibliography}{00}
\bibitem{footnote} Eq.~(\ref{convex}) is not violated if the concurrence in the symmetric state is zero.
\end{thebibliography}
\end{document}